%% file: master.tex
\documentclass[krantz2,ChapterTOCs]{krantz}

\usepackage[utf8]{inputenc}   
\usepackage[T1]{fontenc}      
\usepackage{lmodern}          
\usepackage{graphicx}         

\usepackage{amsmath}          
\usepackage{amsfonts,amssymb} 
\usepackage{mathtools}        

\usepackage{subcaption}       
\newcommand{\subfloat}[2][need a sub-caption]{
  \hspace*{\fill}\subcaptionbox{#1}{#2}\hspace*{\fill}}
\usepackage{booktabs}         

\usepackage{url}
\usepackage{natbib}
\usepackage[sectionbib]{bibunits}
\usepackage{makeidx}

\usepackage[disable]{todonotes}

\let\code=\texttt
\newcommand{\pkg}[1]{\code{#1}}
\newcommand{\abk}[1]{\mbox{#1}.\ }
   
\newcommand{\ie}{\abk{\textit{i.\,e}}}

\renewcommand{\P}{\operatorname{\mathsf{Pr}}}
\newcommand{\abs}[1]{\lvert#1\rvert}
\newcommand{\ind}{\mathbf{1}}

\makeindex

\begin{document}

\bibliographyunit[\chapter]
\defaultbibliographystyle{apalike}
\renewcommand{\bibname}{References}

\frontmatter

\title{Handbook of Infectious Disease Data Analysis}  
\author{Leonhard Held, Niel Hens, Philip O'Neill, Jacco Wallinga}

\cleardoublepage
\setcounter{page}{7} 
\makeatletter\gdef\chapterauthor{\@caplusone}\makeatother


\mainmatter

\setcounter{part}{4}

\setcounter{chapter}{5}

\include{forecasting}

\printindex

\end{document}

%% file: forecasting.tex
\chapterauthor{Leonhard Held}{%
  University of Zurich 
}
\chapterauthor{Sebastian Meyer}{%
  Friedrich-Alexander-Universit\"at Erlangen-N\"urnberg 
}

\chapter{Forecasting Based on Surveillance Data}

\section{Introduction}
Epidemic modelling has at least three distinct aims: Understanding the
spread of infectious diseases, identifying suitable measures to
control the spread of an epidemic, for example through isolation or
vaccination \citep[Section 1.5]{DaleyGani}, and predicting the future
course of an epidemic. Mathematical models are often used to
better understand the dynamics of infectious
disease spread and the effects of control measures \citep[Section
1.5]{Keeling.Rohani2008}, but are less oriented towards
predictions.  In recent years, more emphasis has been placed on
the development of predictive models and methods. The goal of
this chapter is to review the literature in this area and to describe how general
statistical principles from the forecasting literature can be applied
to evaluate the quality of epidemic forecasts. The described methods
will also be illustrated in two case studies, where we assess competing
approaches to forecast time series of infectious disease counts.

\paragraph{The history of forecasting epidemics}

Predicting the future course of epidemics has
been a desire of mankind for a long time. Scientific forecasting
based on mathematical models dates back to the pioneering work
by \citet{Baroyan_etal1971}, who predicted the
course of an influenza epidemic for the main cities in the Soviet
Union. The rise of new infectious diseases has been a
major trigger of novel forecasting methods, such as for
AIDS \citep[Section~6.2]{DaleyGani} and SARS \citep{hufnagel:2004}.
Furthermore, meteorologic forecasting methods have been adopted in
epidemiological research, including
modelling \citep{viboud_etal2003,shaman.karspeck2012}
and assessment \citep{paul_held2011,held.etal2017}
techniques. Recent developments in influenza forecasting have focussed on
the integration of
search logs from Google \citep{dukic.etal2012,shaman.karspeck2012,Yang_etal2015},
social media data from Twitter \citep{paul.etal2014},
combinations thereof \citep{santillana.etal2015},
Wikipedia article views \citep{generous.etal2014,hickmann.etal2015},
or human mobility data \citep{pei.etal2018}.
Compared to weather forecasting, epidemic forecasting is still in its
infancy, and the human component makes it particularly
challenging \citep{who2014,moran.etal2016}.


\paragraph{Forecasting competitions}

In order to develop better epidemic forecasting methods,
the World Health Organization has joined forces in an informal
consultation with more than 130 global experts \citep{who2016},
and the Centers for Disease Control and Prevention in the USA have
organized several seasonal forecasting
competitions \citep{Biggerstaff2016,BIGGERSTAFF2018}.
Real-time influenza forecasts are now
provided by Nicholas Reich and co-workers at the
ReichLab \citep{Tushar_etal2017}, see \url{http://reichlab.io/flusight/}.
Other recent competitions include the
DAPRA challenge on forecasting chikungunya \citep{DelValle2018},
the White House/NOAA challenge on forecasting dengue \citep{Buczak2018},
and the RAPIDD ebola forecasting challenge \citep{VIBOUD201813}.

\paragraph{Forecasting targets}

Several quantities are of interest in epidemic forecasting, such as
timing of and incidence in the peak week \citep{ray.etal2017},
onset week \citep{pei.etal2018},
cumulative incidence \citep{LegaBrown2016},
weekly incidence \citep{paul_held2011,Reich2016},
outbreak size and duration \citep{farrington-etal-2003},
and the epidemic curve \citep{JIANG200990}.
Also of public health relevance are stratified forecasts of the above
quantities, for example by region or by age group \citep{held.etal2017}.
Forecasting targets for seasonal influenza epidemics in particular have
been reviewed previously \citep{chretien.etal2014,nsoesie.etal2014}.
More recently, the weekly proportion of doctor visits due to
influenza-like illness is becoming a popular forecasting
target \citep{BIGGERSTAFF2018}.


\section{Evaluating point forecasts}
\label{sec:pointForecasts}

A point forecast is usually made for a
continuous or integer (often count) outcome, for example the disease
incidence in the next week, the number of newly confirmed cases in a
certain time interval, or the timing of the peak week of an epidemic.
Several measures have been proposed and used to evaluate the quality
of a point forecast.
\citet{gneiting2011} gives a comprehensive discussion of suitable
measures for point predictions.

To introduce some notation, let $\hat y_i$ denote the point forecasts
of the observations $y_i$, $i=1, \ldots, n$. Given a suitable
non-negative scoring function $S(\hat y_i, y_i)$,
the overall predictive performance can be assessed with the mean score
\[
\bar S = \frac{1}{n} \sum_{i=1}^n S(\hat y_i, y_i).
\]
Scoring functions are usually negatively oriented, so the smaller a
score, the better the forecast. Commonly used scoring functions are
the {\em absolute error} (AE) $S(\hat y, y)=\abs{\hat y - y}$, the
{\em squared error} (SE) $S(\hat y, y)=(\hat y - y)^2$, the {\em
  absolute percentage error} $S(\hat y, y)=\abs{\hat y - y}/y$ and the
{\em relative error} $S(\hat y, y)=\abs{\hat y - y}/\hat y$. Note that
the latter two scoring functions require $y$ and $\hat y$ to be
positive, respectively.  Other summary measures used in the infectious
disease literature may be based on one of the scoring functions mentioned
above, for example the {\em root mean squared error} (RMSE) or the {\em
  relative mean absolute error}
\citep{HYNDMAN2006679,reich.etal2016}. The latter is defined as the
ratio of the {\em mean absolute errors} of two competing forecasts and is not
to be confused with the {\em mean relative absolute error} of one forecasting method.

The small simulation study reported by \citet{gneiting2011}
reveals the key differences of the four scoring functions listed above. It is
based on a time series model commonly used in econometrics, but the
results equally apply to other fields.
A simple conditionally
heteroscedastic Gaussian time series model is used to generate a
non-negative time series $y_1, \ldots, y_n$. Model-based
one-step-ahead point forecasts (based on the mean of the forecast
distribution) are then compared to naive approaches, such as the so-called
``fence-sitter forecast'', a forecast with constant
predictions. Quite surprisingly, one of the naive methods outperforms
the model-based one-step-ahead forecasts both under the {\em absolute error}
and the {\em absolute percentage error} scoring functions.

\citet{gneiting2011} points out that the model-based forecast is
only optimal under squared error loss but not necessarily for other
loss functions, where other functions of the predictive distribution
will be optimal.  A scoring function is called consistent for a loss
function, if the expected score is minimized when following this loss
function.  
The {\em squared error} scoring function is consistent for the mean,
and the {\em absolute error} scoring function for the median, both
standard loss functions. However, the {\em mean absolute percentage error}
commonly used in influenza forecasting \citep[Table~4]{chretien.etal2014},
is consistent for a nonstandard and rather exotic functional.
To quote \citet{gneiting2011}, ``it thus seems prudent
that authors using this functional consider the intended or unintended
consequences and reassess its suitability as a scoring function.''
Some scoring functions may be problematic {\em per se}, such as the
commonly used correlation coefficient between predictions and
observations. For example, point predictions always twice as
large as the observations are obviously inappropriate, but the
correlation coefficient between predictions and observations will be one.

\section{Evaluating probabilistic forecasts}
In many areas of science, researchers argue that forecasts
should be probabilistic \citep{gneiting.katzfuss2014} and this plea has
recently been taken up in the infectious disease literature
\citep{held.etal2017,Funk177451}.

Evaluating probabilistic forecasts
requires suitable scores to quantify the ``distance'' between a
cumulative distribution function $F$, the forecast, and the observation $y$
that later realizes. Proper scoring rules play a key role, as they
allow for a fair comparison of different probabilistic forecasting
methods.
In addition, the notions of calibration and sharpness are
important. Calibration is a property of both the forecast and the
observations, whereas sharpness is a property of the forecasts
only. The paradigm of probabilistic forecasting is to ``maximize
sharpness subject to calibration'' \citep{gneiting-raftery-2007}.

Much of the literature on probabilistic forecasting focusses on
continuous forecast distributions. However, in infectious disease
epidemiology, the quantities to be predicted are often (incidence or
prevalence) counts, which require suitable extensions of the assessment
techniques \citep{czado-etal-2009,held.etal2017}.

\subsection{Calibration and sharpness}
Calibration is defined as the statistical consistency of probabilistic
forecasts and the observations
\citep{GneitingBalabdaouiRaftery2007}. For continuous outcomes,
calibration is usually assessed with the probability integral
transform (PIT). Specifically, the PIT is $F(y)$, so will be uniformly
distributed if the observation $y$ is a realisation from the forecast
$F$.  In practice we will compute PIT values for a series of forecasts
and visually assess their distribution in a histogram. Modifications
for count data exist \citep{czado-etal-2009}.

For binary data, the
calibration slope studies the association between the
(logit-transformed) predicted probabilities with the binary outcomes
in a logistic regression \citep{cox1958,Steyerberg2009}. Calibration
can also be visually assessed in a calibration curve
\citep{gneiting.katzfuss2014}.

Statistical calibration tests are
commonly employed to assess the evidence for miscalibration. Different
methods exist for continuous outcomes
\citep{mason.etal2007,held.etal2010}, count outcomes
\citep{wei.held2014} and binary outcomes \citep{spiegelhalter1986}, but
they may require certain distributional assumptions on the forecasts.

For continuous forecasts, sharpness is usually defined as the width of
the associated prediction intervals. For multivariate forecasts,
sharpness is defined based on the predictive covariance
matrix, for example its determinant \citep{gneiting-etal-2008}.

\subsection{Proper scoring rules}
Proper scoring rules are summary measures of the predictive
performance of probabilistic predictions $Y_i \sim F_i$,
$i=1,\ldots,n$, allowing for a joint assessment of calibration and
sharpness \citep{gneiting.katzfuss2014}. As in
Section~\ref{sec:pointForecasts} for point forecasts, the mean score
across forecasts is usually reported and compared across different
forecasting methods.  For ease of presentation we drop the index $i$
in this section and compare the probabilistic forecast $Y \sim F$ with
the actual observation $y$.

The definition of propriety is
mathematically somewhat demanding and we refer the interested reader
to the review paper by \citet{gneiting-raftery-2007}.
Put simply, a proper scoring rule
ensures that a forecaster reports his actual forecast and does not
obtain a better score in expectation by digressing from his
belief. Here we take scores to be negatively oriented penalties that
forecasters wish to minimize, so the smaller a score, the better the
forecast.

For forecasts of binary data $y \in \{0, 1\}$, commonly used scores are
the {\em logarithmic score} (LS), \ie the negative log-likelihood of the
observation under the forecast distribution,
and the {\em Brier score} (BS), also known as probability score.
Specifically, let $\pi$ denote the predicted
probability of the observation $y=1$, then
\begin{eqnarray*}
\mbox{LS}(F, y) & = & - y \log(\pi) - (1-y) \log(1 -\pi), \mbox{ and}\\
\mbox{BS}(F, y) & = & (\pi-y)^2. \\
\end{eqnarray*}
Both scores are proper, whereas the {\em absolute score}
$\mbox{AS}(F, y) = \abs{\pi-y}$ can be shown to be improper
\citep[Chapter~9]{Held.SabanesBove2013}, so should not be used.

A probabilistic
forecast for count data $y \in \{0, 1, 2, \ldots\}$ is represented
by probabilities $\pi_{k} = \P(y=k)$, $k \in \{0, 1, 2, \ldots\}$,
where $\sum_{k=1}^\infty \pi_{k}=1$. The logarithmic score for the
observation $y=j$ now simply
reads $\mbox{LS} = \log(\pi_{j})$ where $\pi_{j}$ is the probability
of the observation $y=j$.
An extension of the Brier score to forecasts of count data
is the {\em ranked probability score}
\[
\mbox{RPS}(F, y) =  \sum_{j=1}^\infty (P_{j} - \ind\{y \leq j\})^2.
\]
where $P_{j} = \sum_{k=0}^j \pi_{k}$ and $\ind$ denotes the indicator function.
Probabilistic count forecasts
are often summarized with the mean $\mu$ and the variance
$\sigma^2$. A proper score that only uses these first two moments is
the {\em Dawid-Sebastiani score}
\[
\mbox{DSS}(F, y) =  2 \log \sigma - (y - \mu)^2/\sigma^2.
\]
This score can easily be generalized to multivariate predictions,
\begin{equation}
\label{eq:mdss}
\mbox{mDSS}(F, \boldsymbol{y}) = \log \abs{\boldsymbol{\Sigma}} +
(\boldsymbol{y}-\boldsymbol{\mu})^\top\boldsymbol{\Sigma}^{-1}(\boldsymbol{y}-\boldsymbol{\mu}),
\end{equation}
which depends only on the mean vector $\boldsymbol{\mu}$ and the
covariance matrix $\boldsymbol{\Sigma}$ of the predictive
distribution. The first term $\log \abs{\boldsymbol{\Sigma}}$ in \eqref{eq:mdss} is called the log
determinant sharpness (logDS) and is
recommended as a (multivariate) measure of
sharpness \citep{gneiting-etal-2008}. To avoid unnecessarily large numbers it is
common practice to report scaled versions of logDS and mDSS,
obtained through division by $2d$ where $d$ is
the dimension of the observation $\boldsymbol{y}$. A possible alternative
to mDSS is the {\em energy score}, a multivariate
extension of the ranked probability score
\citep{gneiting-etal-2008,held.etal2017}.

The question arises which scoring rule should be used in practice.
The logarithmic score is known to be sensitive to outliers if $\pi_{j}
= 0$ is close to zero. The ranked probability score is reported to be
less sensitive to extreme observations \citep{gneiting-raftery-2007}
and provides an attractive alternative. The DSS is particularly useful
if the first two predictive moments are available but not the whole
predictive distribution. It has also been argued that the choice of a
scoring rule should take into account the costs of bad forecasts
\citep{MR3150456}. In practice we recommend to evaluate several scores to
obtain a more robust comparison of predictive performance.

The computation of the different scores depends on whether the forecast
distributions are known analytically or derived from simulations.
Care has to be taken with simulation-based forecasts
as scores can become numerically instable, for
example the logarithmic score for count forecasts will be infinite if
an observation $y=j$ has empirical frequency equal to zero in the
forecast samples.
A possible remedy is to apply a
Rao-Blackwell/importance sampling scheme \citep{GelfandSmith1990} and
to average the conditional predictive distributions, see also
\citet[Supp Mat Section~3.2]{ray.etal2017}. Alternatively one may use
kernel density estimation as
implemented in the \textsf{R}\nocite{R:base}~package
\pkg{scoringRules} \citep{jordan.etal2018}.
The Dawid-Sebastiani score for
multivariate forecasts can be numerically unstable if the predictive
covariance matrix $\boldsymbol{\Sigma}$ is estimated from the empirical covariance
matrix of Monte Carlo samples \citep{scheuerer.hamill2015}.

For independent forecasts, a simple paired $t$-test or a permutation
test can then be used to assess the evidence for differences in
predictive performance between two competing forecasting methods.  If
the forecasts are dependent as in a series of sequential $k$-step
ahead forecasts, the Diebold-Mariano test can be used to account
for the correlation between scores
\citep{DieboldMariano1995,gneiting.katzfuss2014}.

\section{Applications}

We now describe two applications, in which we compare the quality of
forecasts provided by different models and methods.
The data and code to reproduce these analyses are available in the
\textsf{R} package \pkg{HIDDA.forecasting} \citep{R:HIDDA.forecasting}
at \url{https://HIDDA.github.io/forecasting}.
The code is presented in several package vignettes,
which also give some additional results.


\subsection{Univariate forecasting of influenza incidence}

\begin{figure}[bt]

{\centering \includegraphics{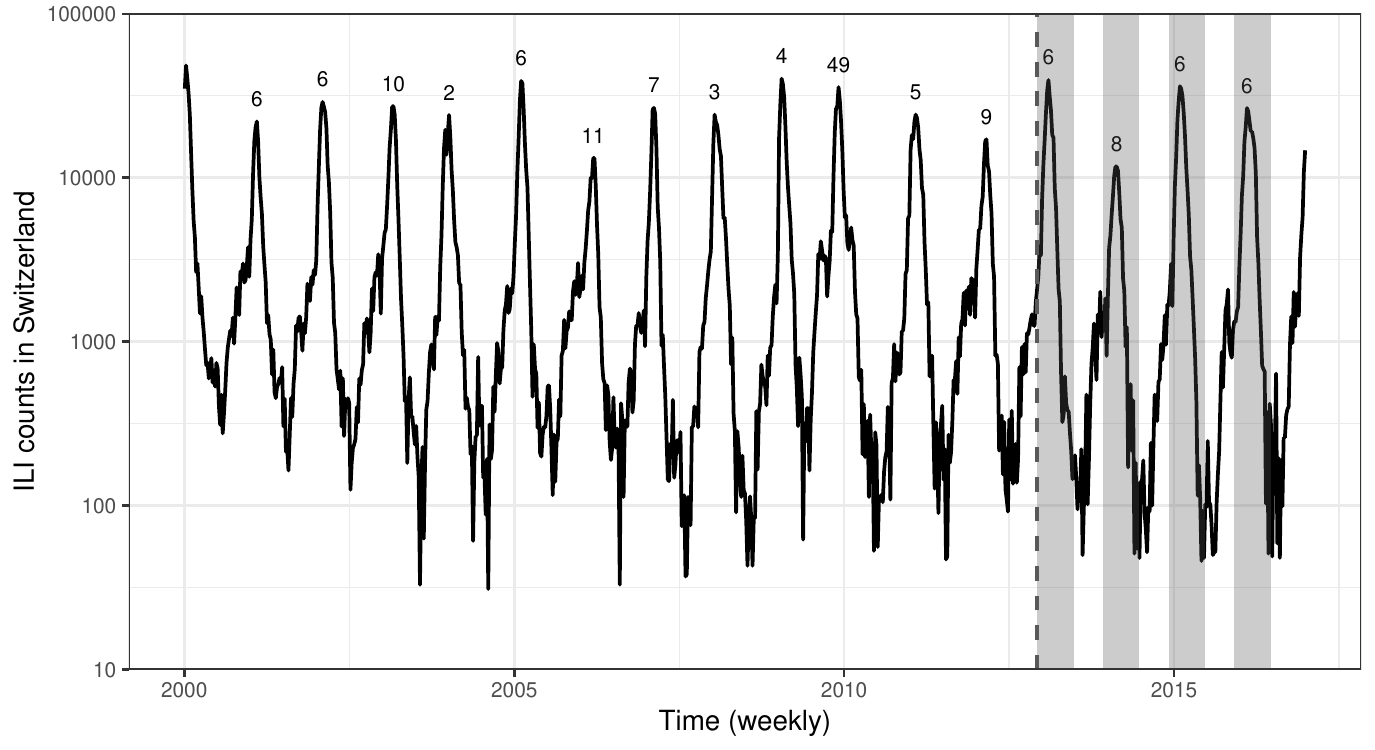} 

}

\caption[Surveillance of influenza-like illness (ILI) in Switzerland, 2000--2016]{Surveillance of influenza-like illness (ILI) in Switzerland, 2000--2016. The last 213 weeks (to the right of the vertical dashed line) are used to assess one-week-ahead forecasts. Long-term forecasts are assessed for the last four seasons starting in the first week of December (shaded time periods). The seasonal peaks are labelled with the corresponding ISO week. Note that the y-axis employs a log-scale for the ILI counts.}\label{fig:CHILI}
\end{figure}

We consider a time series of weekly incidence counts on influenza-like illness
(ILI) in Switzerland from 2000 to 2016 (Figure~\ref{fig:CHILI}).
The last 213 weeks (starting from
2012-12-04) are used to assess
one-week-ahead forecasts. Four sets of 30-weeks-ahead long-term forecasts
(shaded time periods in Figure~\ref{fig:CHILI}) are computed for each of the
last four seasons, conditional upon all data prior to the respective
first week of December.

We have used five different methods to produce forecasts of influenza
activity in Switzerland:
an ARMA(2,2) time series model as estimated by \code{auto.arima()}
from the \pkg{forecast} package \citep{R:forecast},
an observation-driven ARMA model for negative-binomial counts
as implemented in the \pkg{glarma} package \citep{R:glarma},
the endemic-epidemic negative-binomial time-series model \code{hhh4}
from the \pkg{surveillance} package \citep{meyer.etal2017},
Facebook's forecasting tool \pkg{prophet} \citep{R:prophet},
as well as a recently proposed method based on kernel conditional density
estimation \citep{ray.etal2017}, which we implemented following the code
provided at
\url{https://github.com/reichlab/article-disease-pred-with-kcde}.
Note that forecasts from \code{arima} and \code{prophet} are
generated on a log-scale.
All models were configured to account for yearly seasonality.
In the first three models, seasonality was represented by parametric sine-cosine
regressors \citep{held.paul2012} with frequency $2\pi/52.1775$
(derived from the average number of calendar weeks per year).
For \code{prophet} and \code{kcde}, we followed the documented examples
(see the supplementary vignettes for the exact model formulations).
We also included a separate Christmas effect for calendar week 52
(not with \code{kcde}).
For reference, we computed naive historical forecasts for the
213 weeks of the test period, based on fitting
log-normal distributions to the observed counts in the same calendar week
of previous years.

\begin{table}
  \centering
  \caption[Mean scores of one-week-ahead and long-term forecasts of Swiss ILI counts]{%
    Mean scores of the 213 one-week-ahead
    predictions (2012-W49 to 2016-W52) and of the long-term
    forecasts for the last four seasons of the Swiss ILI surveillance data.
    Computing these long-term forecasts with the experimental \code{kcde}
    implementation was too cumbersome. The runtime for computing a
    single one-week-ahead forecast is given in seconds.
    For \code{hhh4} and \code{glarma},
    the long-term results are based on 1000 simulations.}
  \label{tab:OWA+LTF}
  \begin{tabular}{l|rrrr|rrr}
    & \multicolumn{4}{|c|}{One-week-ahead}
    & \multicolumn{3}{|c}{Long-term} \\
Method & RMSE & DSS & LS & runtime & RMSE & DSS & LS \\ 
  \midrule
\code{arima} & 2287 & 13.78 & 7.73 & 0.51 & 8471 & 16.43 & 8.88 \\ 
  \code{glarma} & 2450 & 13.59 & 7.71 & 1.49 & 5558 & 19.61 & 9.12 \\ 
  \code{hhh4} & 1769 & 13.58 & 7.71 & 0.02 & 8749 & 16.13 & 9.25 \\ 
  \code{prophet} & 5614 & 15.00 & 8.03 & 3.01 & 7627 & 16.44 & 8.91 \\ 
  \code{kcde} & 1963 & 13.79 & 7.80 & 1.30 & --- & --- & --- \\ 
  naive & 5010 & 14.90 & 8.06 & 0.00 & 6527 & 15.99 & 8.86 \\ 
  
  \end{tabular}
\end{table}

We computed the RMSE and the mean DSS and LS for the one-week-ahead
forecasts and the long-term forecasts. The results are summarized in
Table~\ref{tab:OWA+LTF}.
As expected, the one-week-ahead foreacasts have smaller (\ie better) mean
scores than the long-term forecasts for all methods considered.
Note that we have not attempted to compute long-term forecasts with
\pkg{kcde} due to excessively long runtimes. The time taken to compute a single
one-week-ahead forecast varied between 0.001 (naive) and 3 (\code{prophet})
seconds.

\begin{figure}[p]
  \centering
  \subfloat[\code{arima}\label{fig:OWA:arima}]{
    \includegraphics[width=0.7\linewidth]{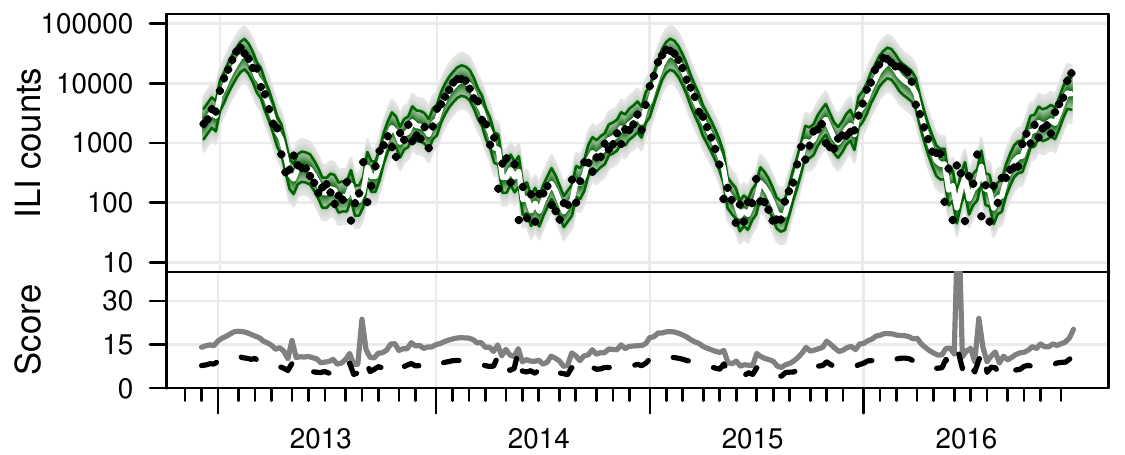}
    \includegraphics[width=0.3\linewidth]{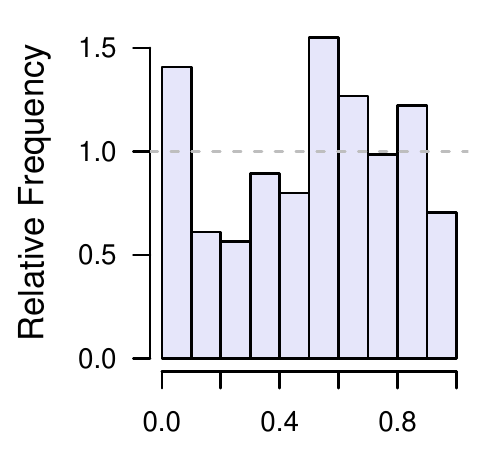}
  }

  \smallskip
  \subfloat[\code{glarma}\label{fig:OWA:glarma}]{
    \includegraphics[width=0.7\linewidth]{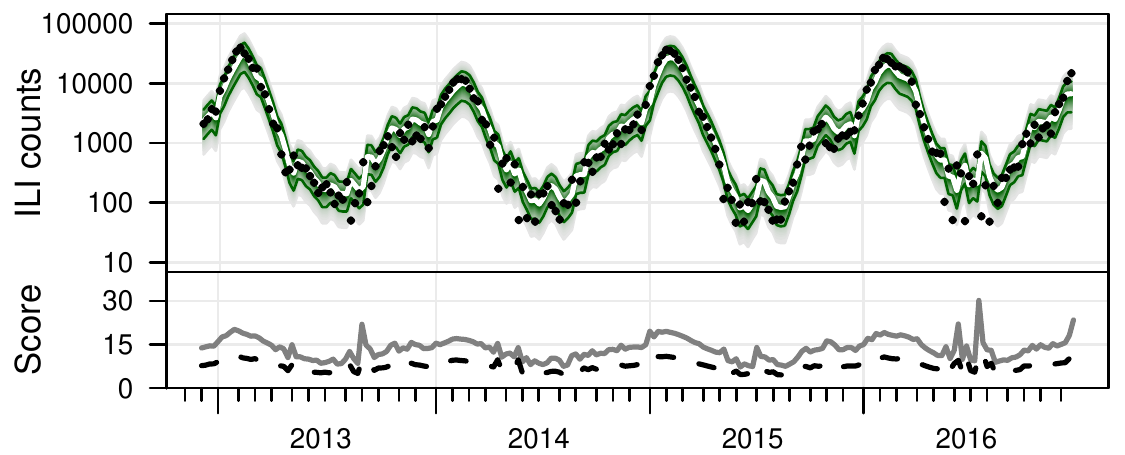}
    \includegraphics[width=0.3\linewidth]{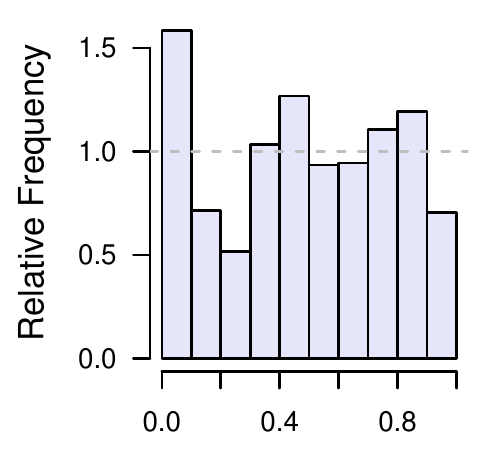}
  }

  \smallskip
  \subfloat[\code{hhh4}\label{fig:OWA:hhh4}]{
    \includegraphics[width=0.7\linewidth]{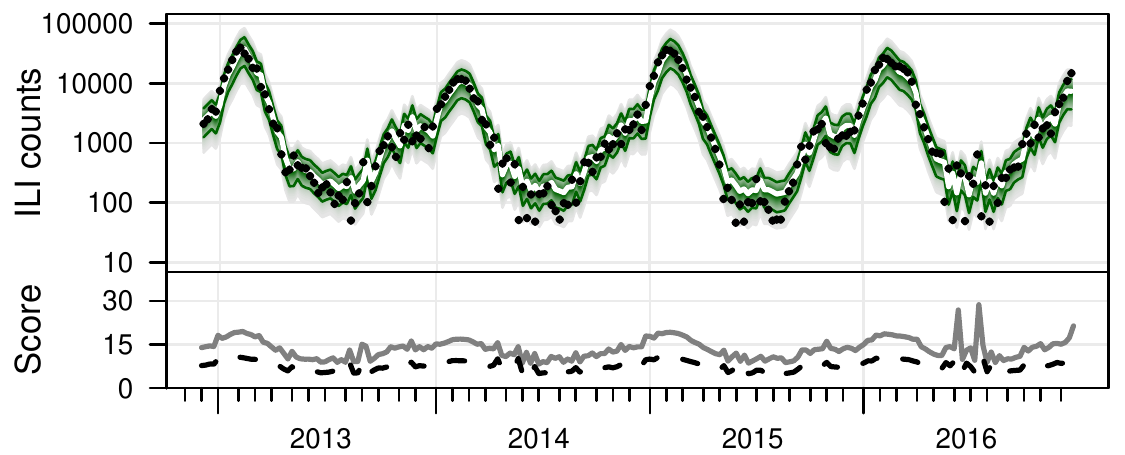}
    \includegraphics[width=0.3\linewidth]{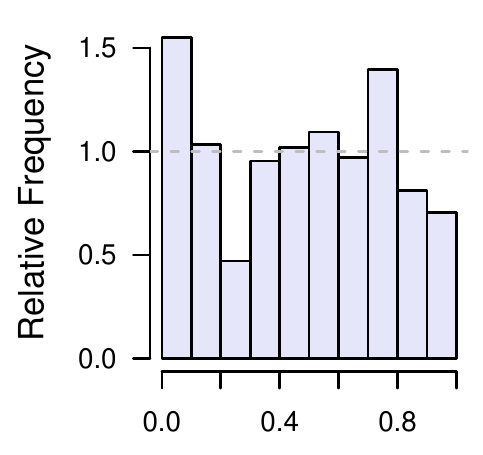}
  }

  \smallskip
  \subfloat[\code{prophet}\label{fig:OWA:prophet}]{
    \includegraphics[width=0.7\linewidth]{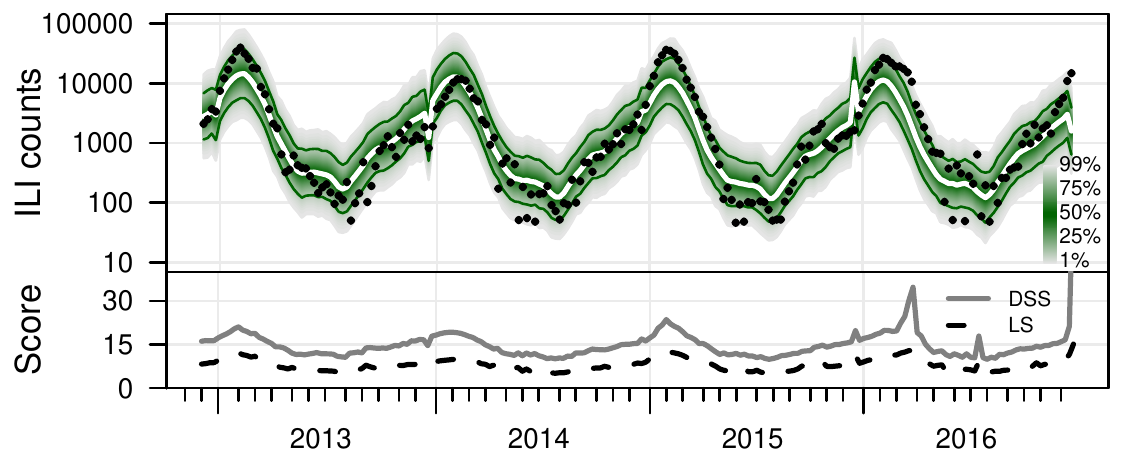}
    \includegraphics[width=0.3\linewidth]{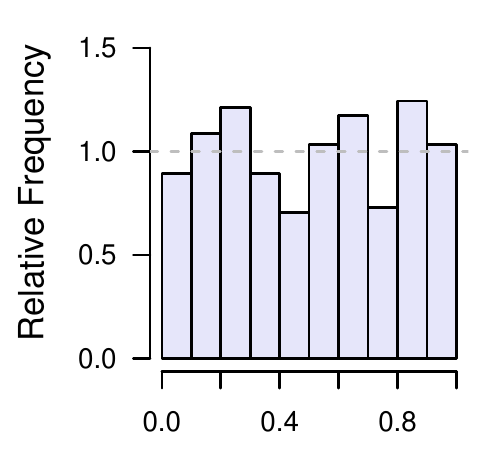}
  }
  \caption[One-week-ahead forecasts of ILI counts in Switzerland]{%
    One-week-ahead forecasts of ILI counts in Switzerland for the last
    213 weeks (2012-W49 to 2016-W52) of
    the available surveillance data. Predictive distributions are
    displayed as fan charts on a log-scale. The 10\% and 90\% quantiles
    and the mean are highlighted. The dots correspond to the
    observed counts and the lower panels show the associated weekly scores.
    The right-hand column shows the corresponding PIT histograms.
    Plots for \code{kcde} and naive forecasts can be found in the
    supplementary vignettes.\label{fig:OWA}}
\end{figure}

The best one-week-ahead forecasts in terms of all scores are obtained with
\code{hhh4} followed by \code{glarma}, whose point forecasts
are slightly worse with an average error of 2450 cases.
The \code{arima} and \code{kcde} methods come next and the worst
one-week-ahead predictions are obtained from \pkg{prophet}, which achieves
similar scores as the naive approach. Figure~\ref{fig:OWA} shows the weekly
forecasts and associated scores as well as the overall PIT histograms,
which are computed based on the method for count data
\citep{czado-etal-2009}.  There is no clear evidence for miscalibration
of any of the one-week-ahead forecasts, but the first three and the naive
(not shown) forecasting methods have a distinct peak of the histogram in the
first bin, indicating some evidence for biased or underdispersed
predictions. The \code{prophet} and \code{kcde} (not shown) forecasts don't have
this problem. PIT histograms for the \code{kcde} and naive forecasts can be
found in the corresponding supplementary vignettes.

\begin{figure}[p]

{\centering \subfloat[\code{arima}\label{fig:LTF_fans1}]{\includegraphics[width=.5\linewidth]{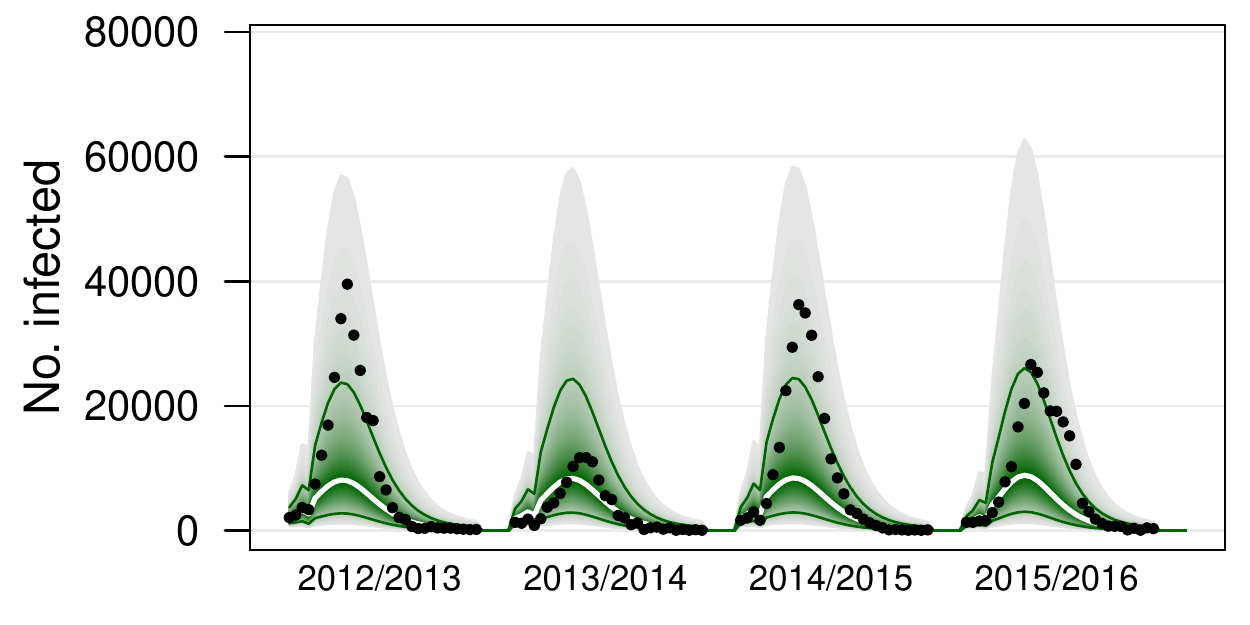} }\subfloat[\code{glarma}\label{fig:LTF_fans2}]{\includegraphics[width=.5\linewidth]{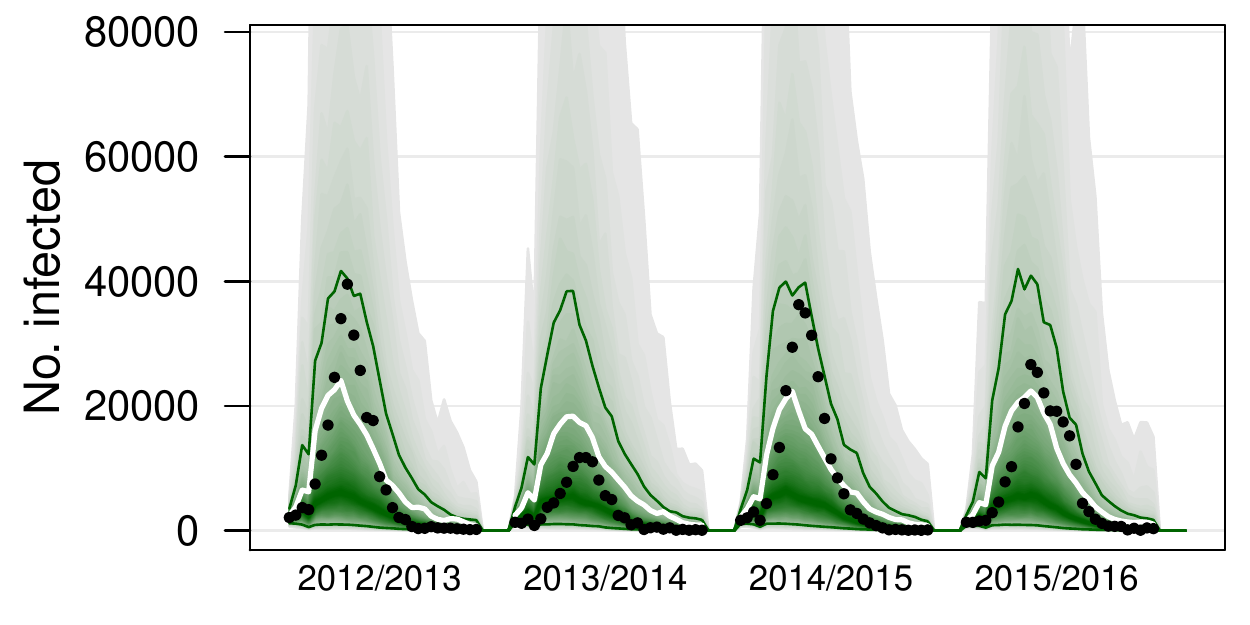} }\newline\subfloat[\code{hhh4}\label{fig:LTF_fans3}]{\includegraphics[width=.5\linewidth]{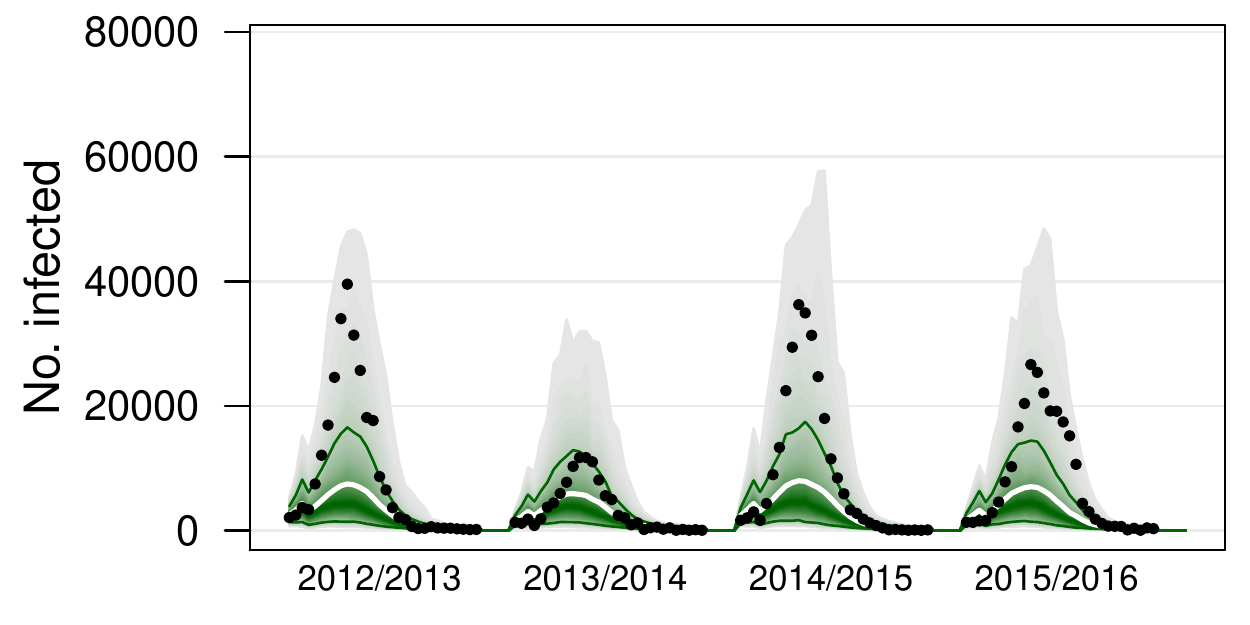} }\subfloat[\code{prophet}\label{fig:LTF_fans4}]{\includegraphics[width=.5\linewidth]{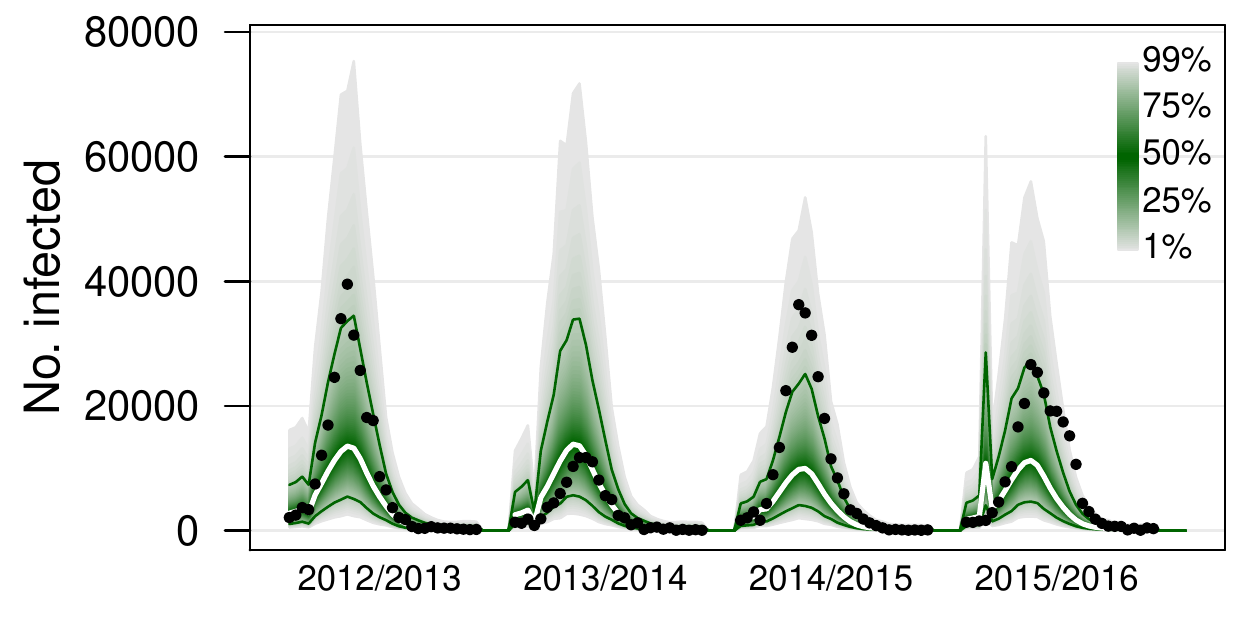} }

}

\caption[Long-term forecasts (30-weeks-ahead) of ILI counts in Switzerland for the last four seasons of the available surveillance data]{Long-term forecasts (30-weeks-ahead) of ILI counts in Switzerland for the last four seasons of the available surveillance data. Predictive distributions are displayed as fan charts, where the 10\% and 90\% quantiles and the mean are highlighted. The dots represent observed counts. In (b), quantiles above 95\% are truncated around the peak weeks with the 99\% quantile reaching seasonal maxima of between 251000 and 325000 cases.}\label{fig:LTF_fans}
\end{figure}

\begin{figure}[p]

{\centering \includegraphics{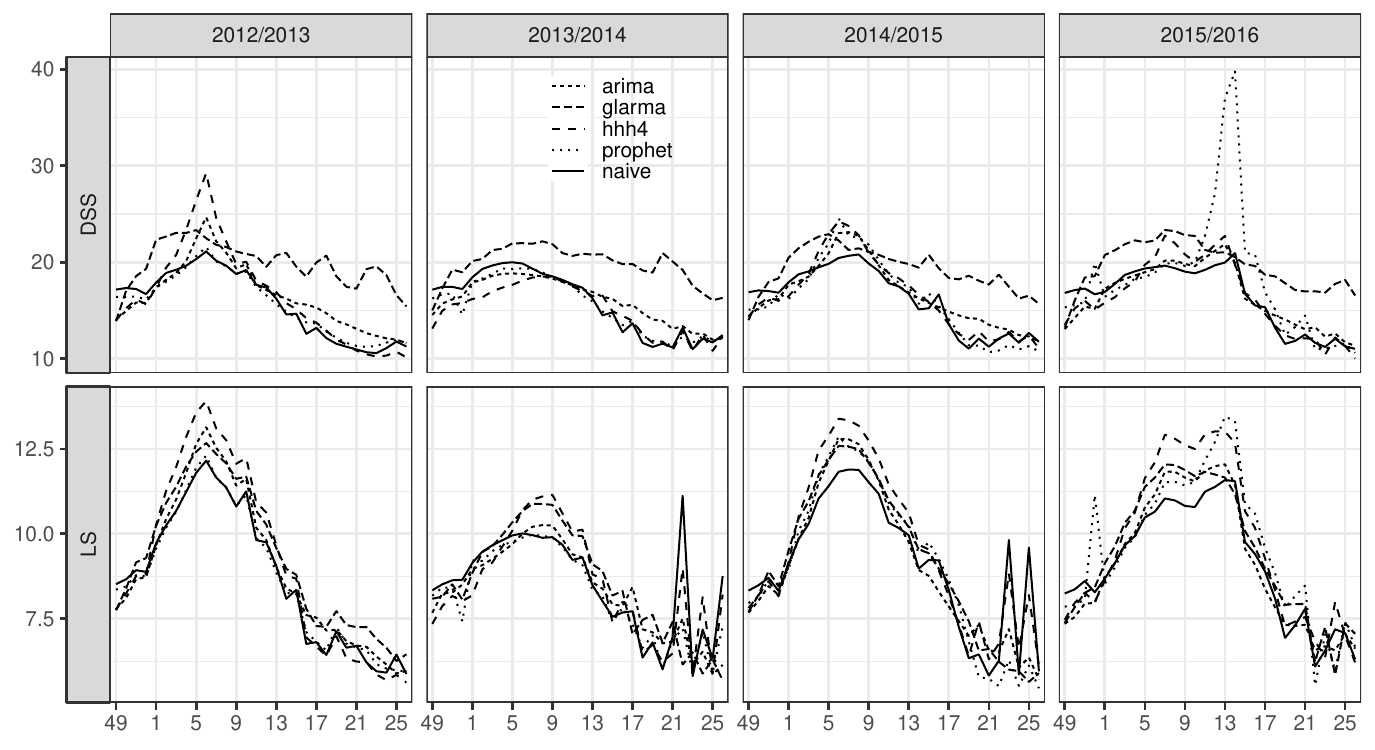} 

}

\caption{Weekly DSS (top) and LS (bottom) values of the different long-term forecasts of Swiss ILI counts displayed in Figure~\ref{fig:LTF_fans} and the naive forecasts.}\label{fig:LTF_scores}
\end{figure}

An interesting result is obtained from the long-term forecasts, where
the best scores are achieved by taking simple historical averages, except for RMSE
where \code{glarma} is better.
For the more sophisticated forecasts shown in Figure
\ref{fig:LTF_fans}, the ranking is less clear.
While \code{hhh4} is the best model in terms of
DSS, \code{arima} is the best model in terms of LS.
Quite surprising is the large DSS value of the \code{glarma} method.
Closer inspection of Figure \ref{fig:LTF_fans} suggests that this is caused
by the excessively large uncertainty of the \code{glarma} long-term
predictions, compared to the other three methods.
From Figure
\ref{fig:LTF_scores} we can see that the \code{prophet} method has a
particularly large DSS score in the 2015/2016 season, where the DSS
score is roughly twice as large as for the other methods. Closer
inspection of Figure \ref{fig:LTF_fans} reveals that this is due to
under-prediction of the incidence combined with an insufficient amount of
uncertainty at the end of the epidemic season.

It may also be of interest to predict the peak week in each of the
four years based on the long-term forecasts. For illustration, we have
computed a probabilistic prediction of the peak week based on the
samples from the \code{hhh4} model only. The median peak week
forecasts is always calendar week 5.
This is quite close to the observed peak weeks 6, 8, 6, and 6, respectively,
shown in Figure~\ref{fig:CHILI}.  However, considerable uncertainty is
attached to these predictions with 2.5\% quantile the second week and
97.5\% quantile week 16 of each season.

\subsection{Multivariate forecasting of norovirus incidence}

\begin{figure}[bht]

{\centering \includegraphics{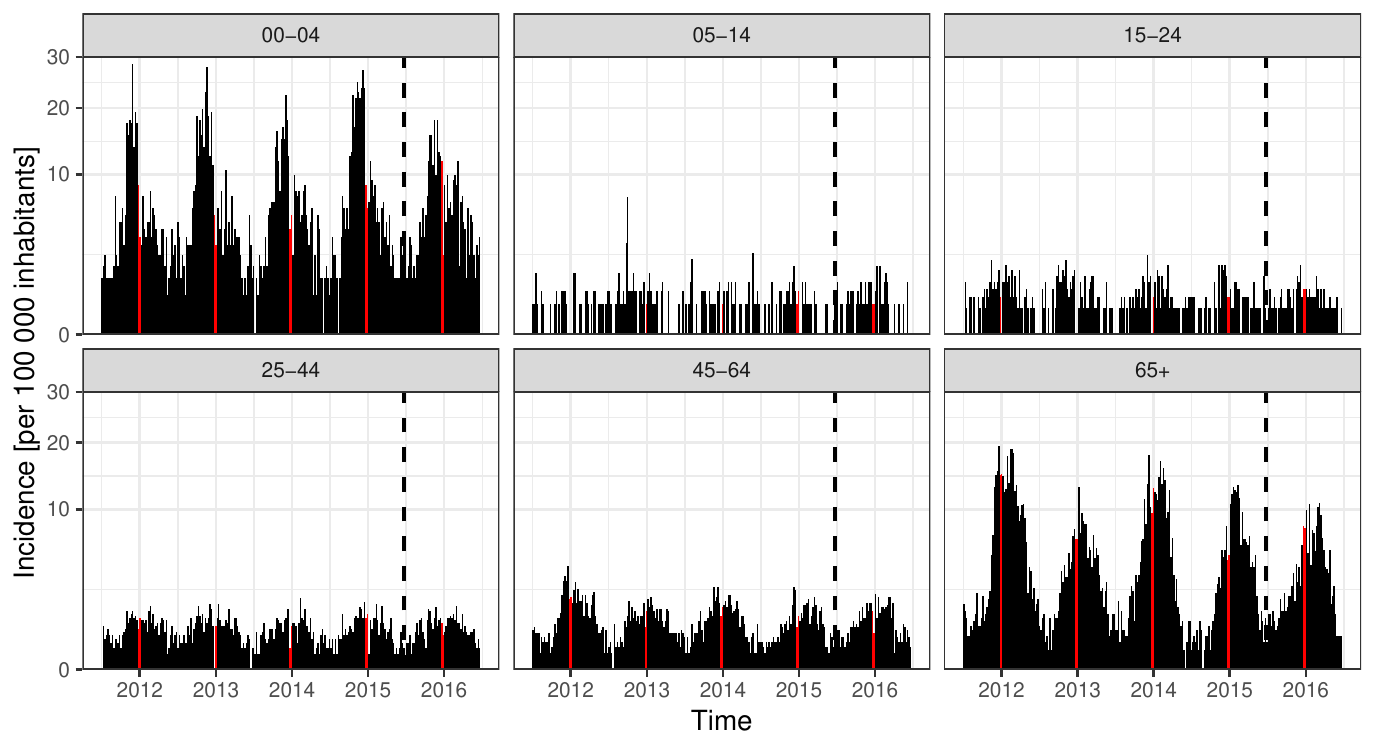} 

}

\caption[Age-stratified incidence of norovirus gastroenteritis in Berlin, Germany]{Age-stratified incidence of norovirus gastroenteritis in Berlin, Germany. The calendar weeks 52 and 1 (Christmas break period) are highlighted. The last 52 weeks (to the right of the vertical dashed line) are used to assess forecasts. Note that the y-axis employs a square-root scale.}\label{fig:BNV}
\end{figure}

To illustrate statistical methods for the assessment of
multivariate forecasts we use age-stratified surveillance data on
norovirus gastroenteritis from Berlin, Germany. We use the
well-established \code{hhh4} modelling framework from the
\pkg{surveillance} package to generate forecasts, building on our
previous analyses of these data \citep{meyer.held2017,held.etal2017}.
Here we only consider models for spatially aggregated counts
$Y_{gt}$, $g=1,\dots,G$,
\ie without additional stratification by city district.
The surveillance data are available from the package
\pkg{hhh4contacts} \citep{R:hhh4contacts}, cover five norovirus seasons from
2011-W27 to 2016-W26, and are stratified into $G=6$ age
groups:
0--4, 5--14, 15--24, 25--44, 45--64, and 65+ years of age.  These
groups reflect distinct social mixing of pre-school vs.\ school
children, and intergenerational mixing.
Figure~\ref{fig:BNV} shows the age-specific incidence time series.
Over the five years, the reported numbers of cases by age group were
2783, 326, 553, 1909, 2530, and 8335, respectively,
corresponding to average yearly incidences of
346 (0--4), 24 (5--14), 30 (15--24), 36 (25--44), 52 (45--64), and 251 (65+)
cases per 100\,000 inhabitants.

\begin{figure}

{\centering \includegraphics{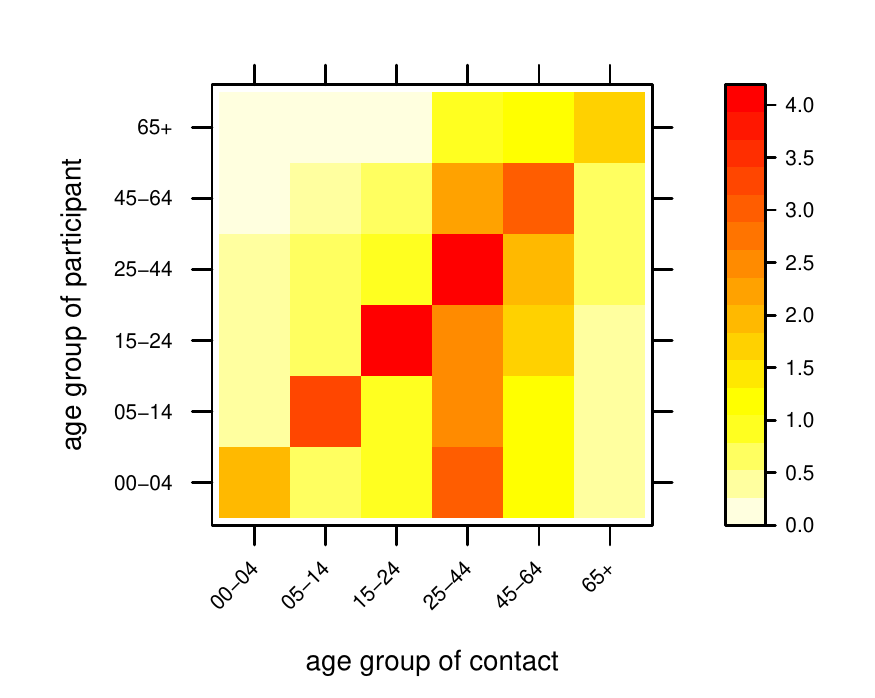} 

}

\caption[Age-structured social contact matrix estimated from the German POLYMOD sample]{Age-structured social contact matrix estimated from the German POLYMOD sample. The entries refer to the mean number of contact persons per participant per day.}\label{fig:C_reci}
\end{figure}

Given the counts from the previous week, $Y_{g',t-1}$, $g'=1, \ldots, G$,
we assume $Y_{gt}$ to follow a negative binomial distribution with a
group-specific overdispersion parameter and mean
\begin{equation*}
\mu_{gt} = \nu_{gt} + \phi_{gt} \sum_{g'} c_{g'g} Y_{g',t-1}.
\end{equation*}
The \emph{endemic} log-linear predictor $\nu_{gt}$ contains group-specific
intercepts, a Christmas break effect (via a simple indicator for the
calendar weeks 52 and 1), and group-specific seasonal effects via
$\sin(\omega t)$ and $\cos(\omega t)$ terms, $\omega=2\pi/52$.
The \emph{epidemic} log-linear predictor $\phi_{gt}$ also contains
group-specific intercepts, but shared seasonality and no Christmas break effect.
The contact weights $c_{g'g}$ are estimated from the German subset of the
POLYMOD survey \citep{mossong.etal2008}, taking into account the reciprocal
nature of contacts \citep{wallinga.etal2006}.
The resulting contact matrix (Figure~\ref{fig:C_reci}) is subsequently
row-normalized to a transition matrix, removing differences in group-specific
overall contact rates.

The model described above is the same as reference model~6
in \citet[Section 3.3]{held.etal2017}.
As alternative models we consider homogeneous mixing
between age groups ($c_{g'g} = 1$), no mixing between age groups
($c_{g'g} = \ind\{g'=g\}$) and a model where we estimate a power
transformation $C^\kappa$ of the contact matrix as described in
\citet{meyer.held2017}.
See the \code{vignette("BNV")} in the supplementary package
\pkg{HIDDA.forecasting} for how to implement these models in \textsf{R}.

\begin{figure}

{\centering \includegraphics{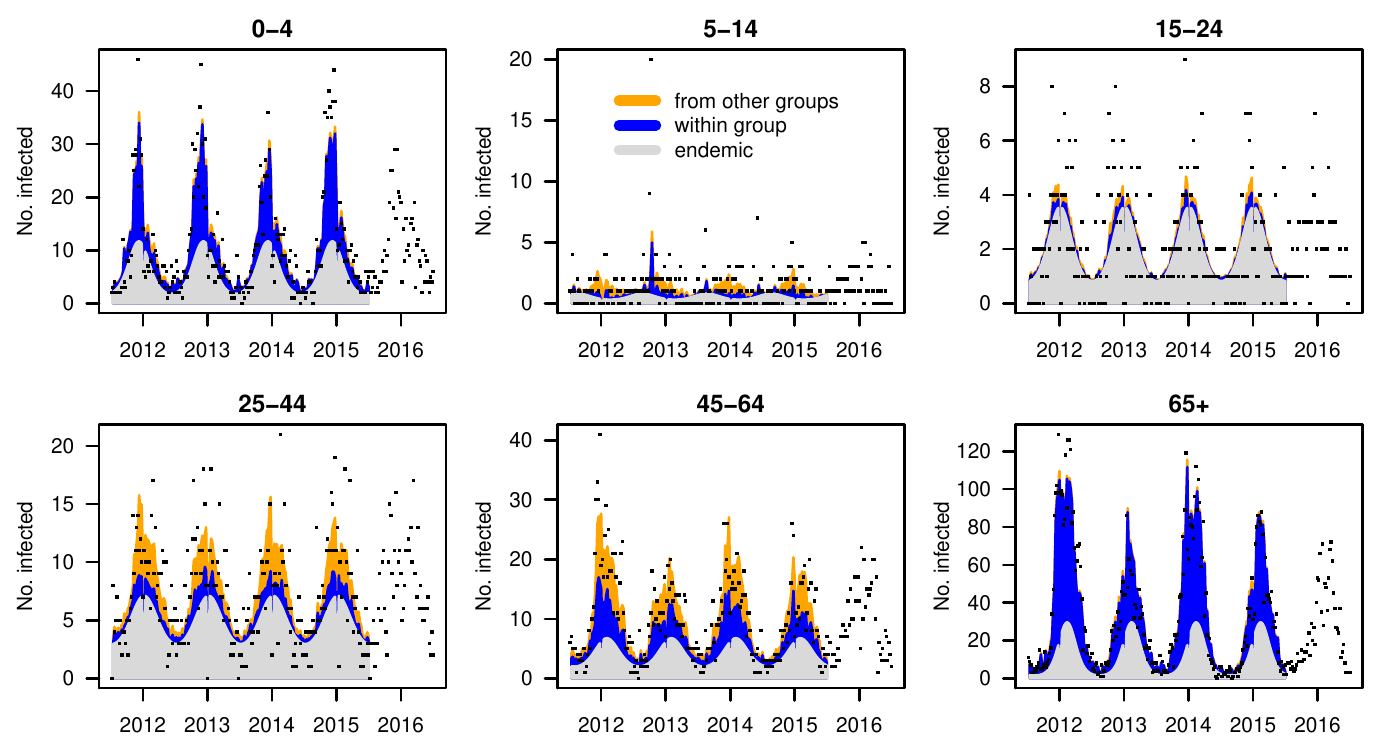} 

}

\caption[Estimated norovirus counts from the AIC-optimal \code{hhh4} model with power-adjusted contact matrix]{Estimated norovirus counts from the AIC-optimal \code{hhh4} model with power-adjusted contact matrix. The disease risk is additively decomposed into endemic and epidemic components. The dots show the observed counts.}\label{fig:BNV_fitted_components}
\end{figure}

We have fitted the models to the first four seasons (2011-W27 to
2015-W26) and evaluated the quality of forecasts
in the subsequent season (2015-W27 to 2016-W26).
Figure~\ref{fig:BNV_fitted_components} shows the fitted
mean components from the AIC-optimal model with power-adjusted contact matrix.
The proportion of the incidence that can be explained by the epidemic component
varies between the age groups.
The endemic-epidemic decomposition actually resembles Figure~4 in
\citet{meyer.held2017} quite well, which is based on a different model,
additionally stratified by city district.
Within the epidemic part, the incorporated social contact matrix results in
predominant within-group reproduction of the disease for the 0--4 and 65+ age
groups, whereas the 25--44 and 45--64 year-old persons inherit a substantial
proportion of cases from other age groups.

\begin{figure}
\captionsetup[subfigure]{labelformat=empty,skip=0pt}
\centering
\subfloat[]{\includegraphics[width=.5\linewidth]{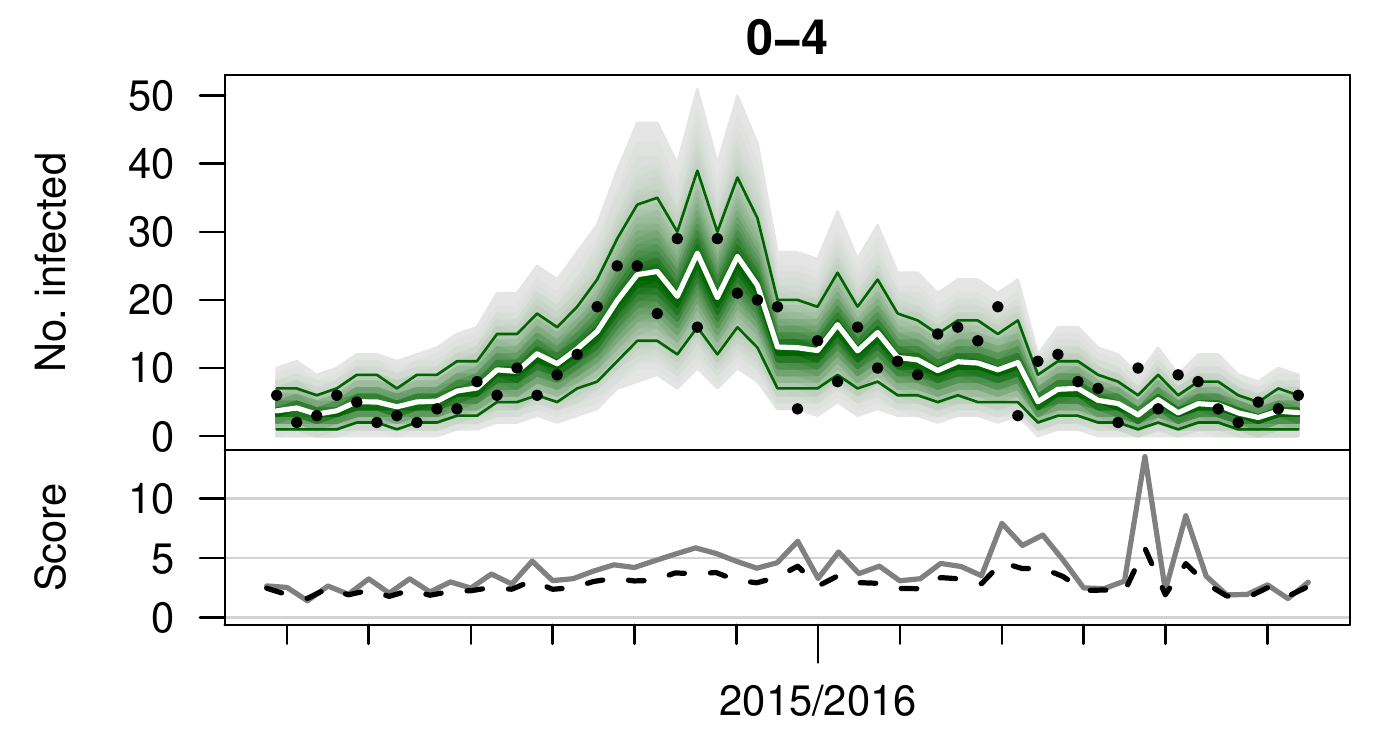}}
\subfloat[]{\includegraphics[width=.5\linewidth]{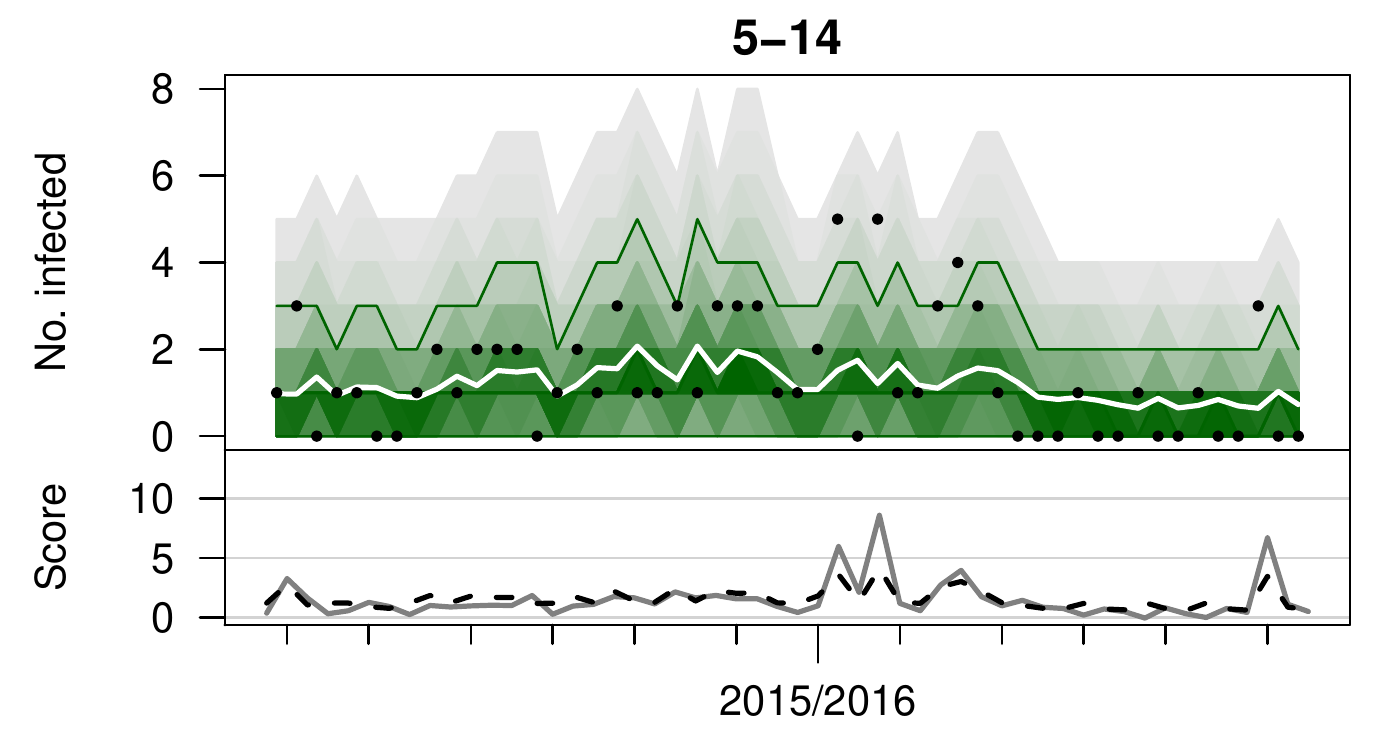}}

\subfloat[]{\includegraphics[width=.5\linewidth]{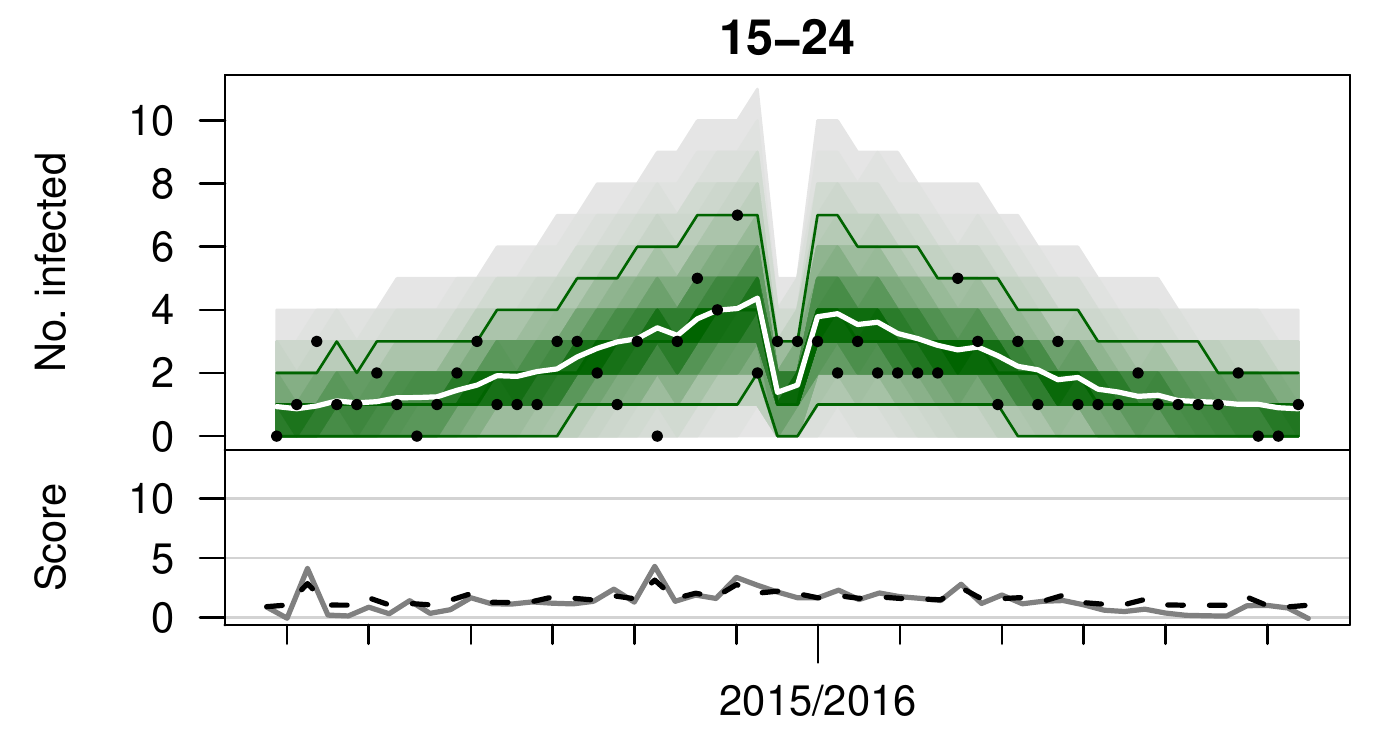}}
\subfloat[]{\includegraphics[width=.5\linewidth]{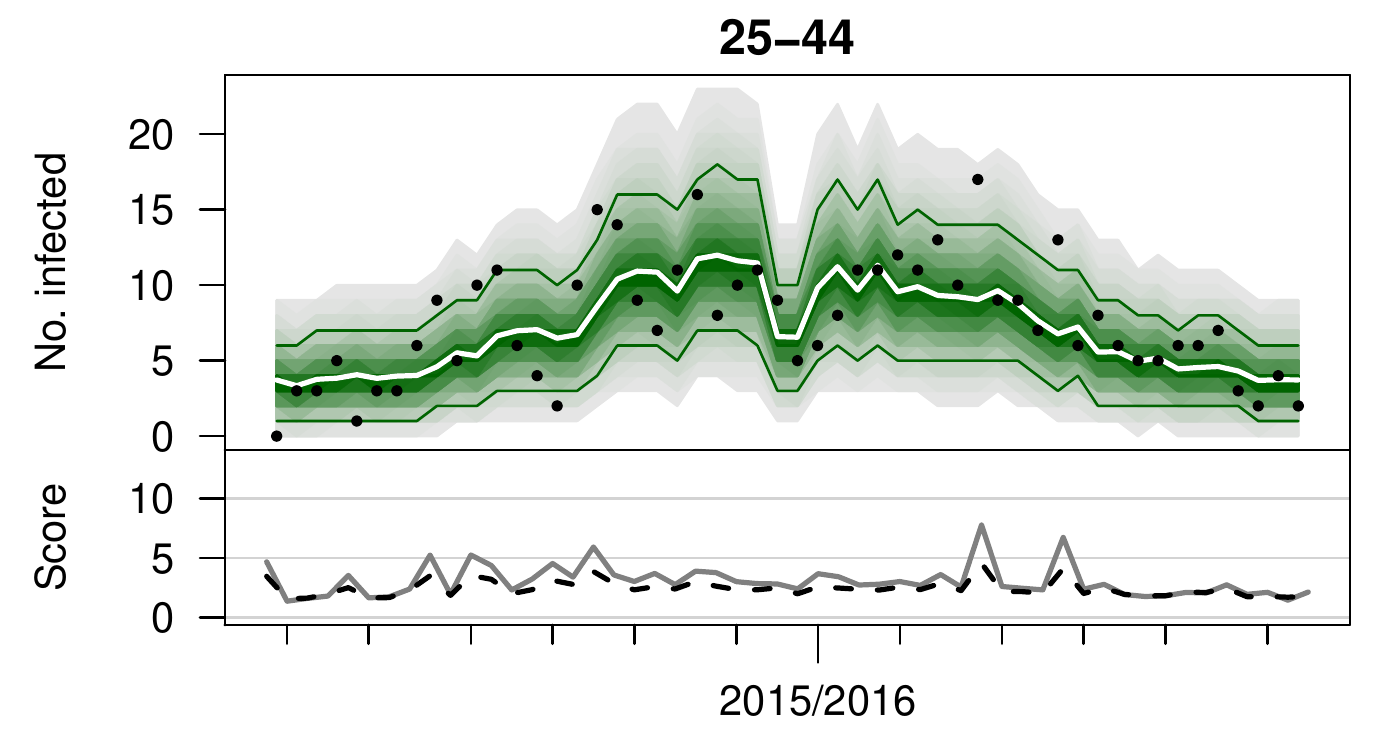}}

\subfloat[]{\includegraphics[width=.5\linewidth]{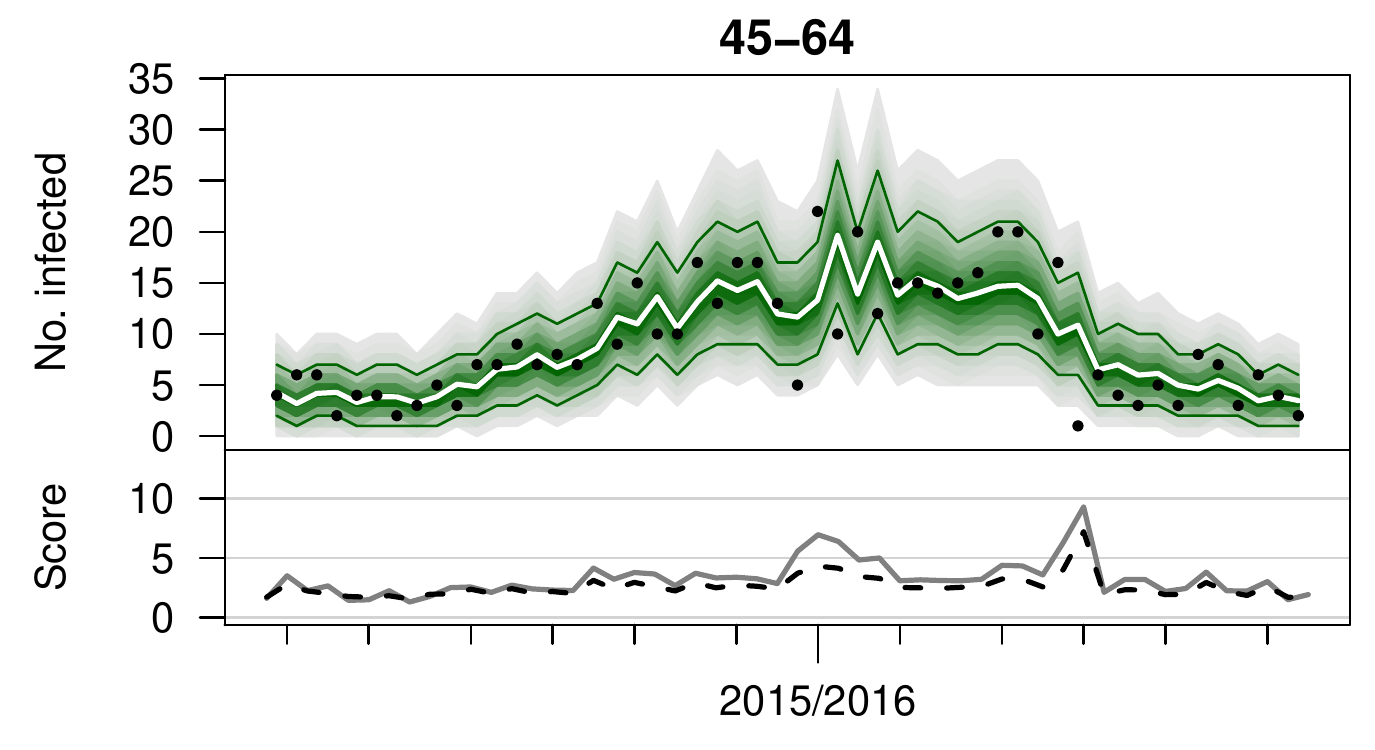}}
\subfloat[]{\includegraphics[width=.5\linewidth]{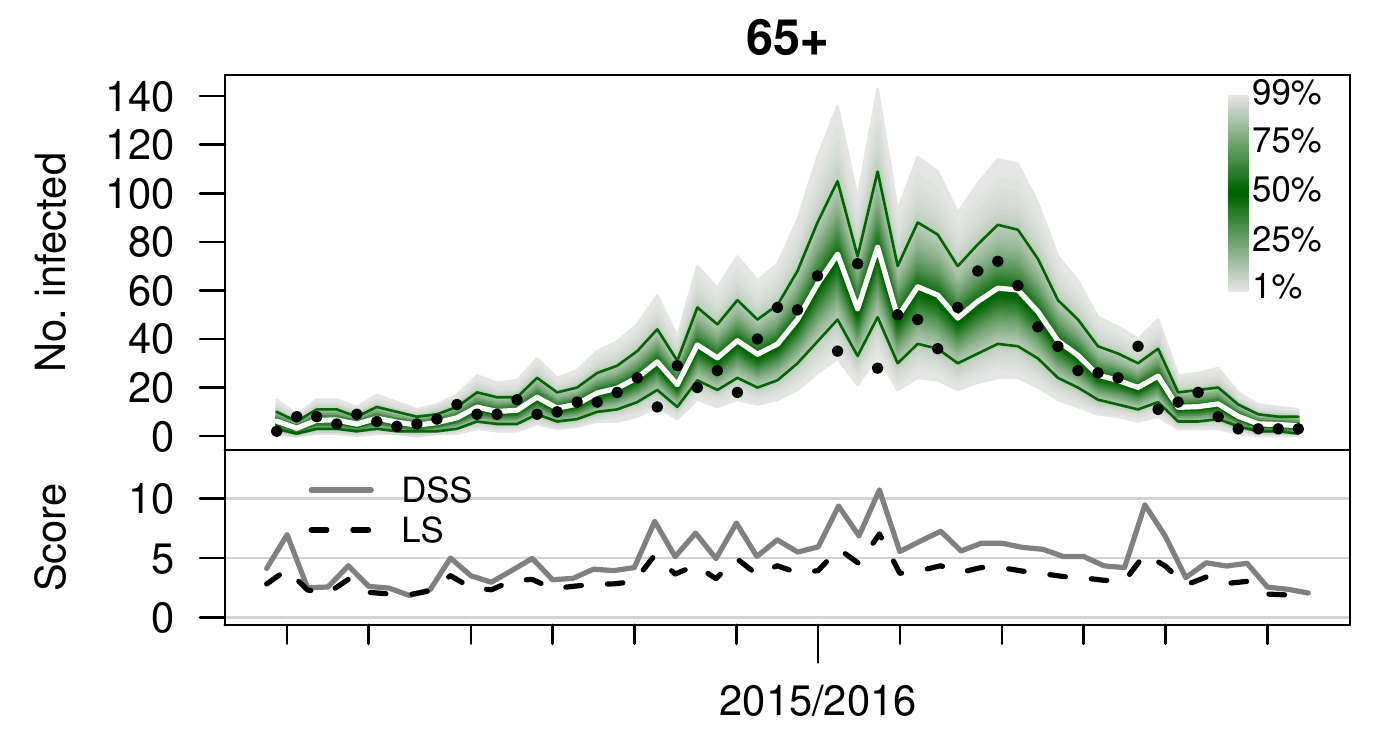}}
\caption[Age-stratified one-week-ahead forecasts of norovirus counts]{%
  Age-stratified one-week-ahead forecasts of norovirus counts in
  Berlin based on the power-adjusted contact model. The 10\% and 90\%
  quantiles and the means are highlighted within the fan charts. The dots
  represent observed counts. The bottom panels show weekly scores.}
\label{fig:BNV_OWA}
\end{figure}

Turning to the predictive performance,
Figure~\ref{fig:BNV_OWA} shows the one-week-ahead forecasts and associated
scores during the last season for the power-adjusted contact model.
Average scores of the different models are
compared in Table~\ref{tab:BNV}.
Although the power-adjusted model gives the
best AIC score on the training data, it is outperformed by the ``no
mixing'' model on the test season, both in terms of DSS and LS. This
can be explained by Table~\ref{tab:GOF_season}, which gives the
log-likelihood contributions of the different models from a separate
fit to the whole time period (training and test). We can see that the
``no mixing'' model has indeed a larger log-likelihood contribution in
2015/16 than all other three models. However, the power-adjusted model
is best in 2011/12, 2013/14 and 2014/15 and second best in 2012/13
(after ``reciprocal''). This explains why the power-adjusted model
performs best in the training, but not in the test period.

\begin{table}
\centering
\caption[Comparison of four \code{hhh4} models for the age-stratified
norovirus time-series]{Comparison of four \code{hhh4} models for the age-stratified
  norovirus time-series from Berlin, assuming different contact matrices.
  The table gives the number of parameters, the AIC for
  the training period, and average scores of the $6
  \times 52$ one-week-ahead forecasts in the last season.}
\label{tab:BNV}
\begin{tabular}{rrrrr}
  & dim & AIC & DSS & LS \\ 
  \midrule
reciprocal & 33 & 6051 & 3.031 & 2.399 \\ 
  homogeneous & 33 & 6132 & 3.093 & 2.420 \\ 
  no mixing & 33 & 6055 & 3.003 & 2.385 \\ 
  power-adjusted & 34 & 6035 & 3.012 & 2.391 \\ 
  \end{tabular}

\end{table}

\begin{table}
\centering
\caption[Seasonal log-likelihood contributions in the norovirus
models from Table~\ref{tab:BNV}]{
  Seasonal log-likelihood contributions in the norovirus
  models from Table~\ref{tab:BNV}, now fitted to the whole time period.
  The last column gives the overall AIC.}
\label{tab:GOF_season}
\begin{tabular}{rrrrrrr}
  & 2011/12 & 2012/13 & 2013/14 & 2014/15 & 2015/16 & AIC \\ 
  \midrule
reciprocal & -747.53 & -758.71 & -763.60 & -726.83 & -740.38 & 7540 \\ 
  homogeneous & -762.44 & -766.59 & -769.68 & -739.53 & -745.16 & 7633 \\ 
  no mixing & -750.36 & -763.29 & -762.37 & -721.71 & -737.47 & 7536 \\ 
  power-adjusted & -744.43 & -759.06 & -762.01 & -721.68 & -738.63 & 7520 \\ 
  \end{tabular}

\end{table}

\begin{figure}[bht]

{\centering \includegraphics{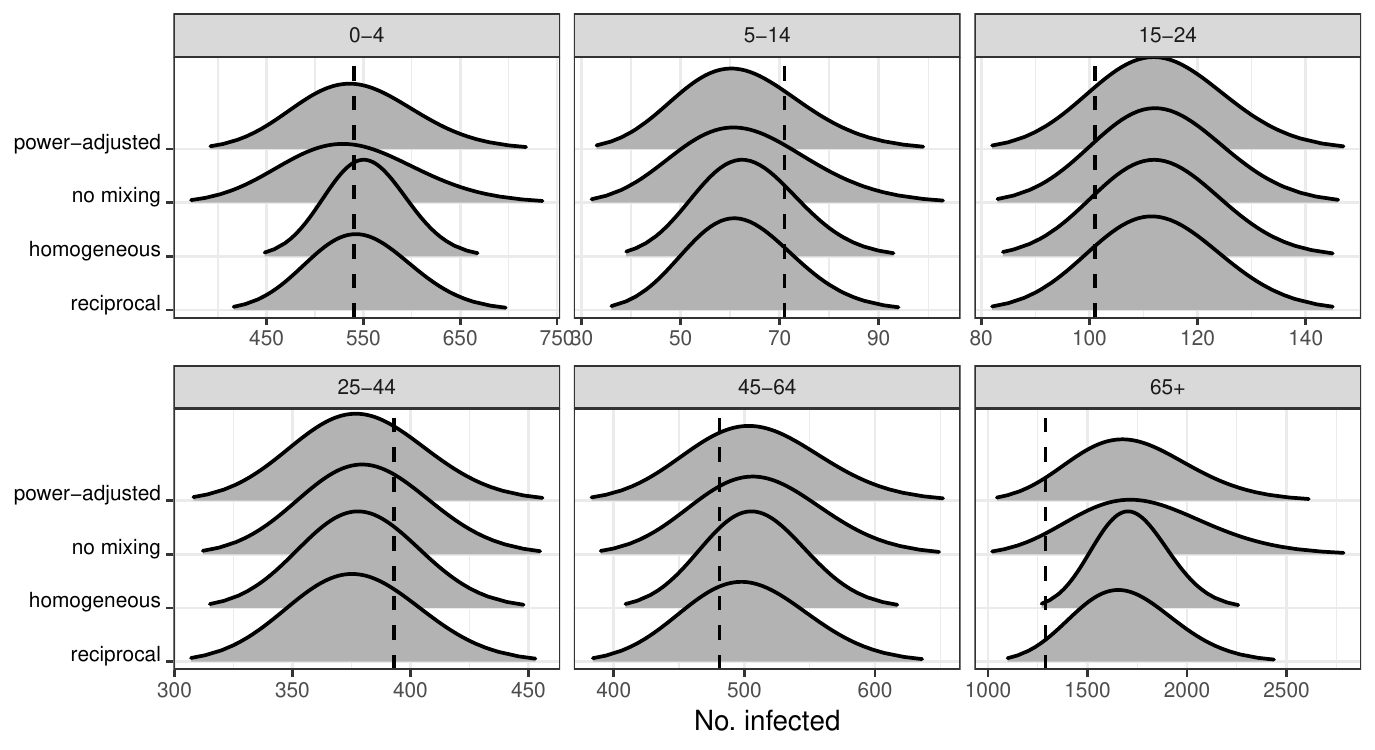} 

}

\caption{Forecast distributions of age-stratified sizes of the 2015/2016 norovirus epidemic from four different models. Shown are negative binomial approximations based on predictive moments, see \citet[Appendix A]{held.etal2017} for details. The dashed vertical lines represent the observed counts.}\label{fig:BNV_LTF_size}
\end{figure}

Multivariate long-term predictions from the four models are
evaluated in Table~\ref{tab:BNV_LTF_DSS} for the test period. The
first two columns give mDSS and logDS for multivariate predictions
(of dimension $6 \times 52$)  across weeks and age groups. The ``no
mixing'' model is again best in terms of mDSS, although both the
``reciprocal'' and the power-adjusted model provide sharper
predictions in terms of logDS.  Predictions of the age-stratified total
number of cases (now of dimension 6) are displayed in
Figure~\ref{fig:BNV_LTF_size}
and the corresponding scores listed in the last two
columns of Table~\ref{tab:BNV_LTF_DSS}. More uncertainty of the
predictions (in age group 65+) is visible for the ``no mixing'' and
power-adjusted models, resulting in larger values of logDS. However,
the increased uncertainty seems to be benificial for the quality of
the forecasts, yealding the smallest mDSS scores for these two
formulations. In contrast, the ``homogeneous'' prediction barely
covers the observed count in age group 65+, resulting in a poor mDSS score.
For comparison, a naive forecaster who independently fits negative binomial
distributions to the age-stratified sizes of the past four seasons would reach
$\mbox{mDSS}=4.084$ (with
$\mbox{logDS}=3.820$),
which is superior to the predictions from the homogeneous contact model.

\begin{table}[bht]
\centering
\caption[Scores of long-term predictions for the last norovirus season]{
  Scaled multivariate Dawid-Sebastiani scores (mDSS) and log determinant
  sharpness (logDS) of long-term predictions for the last norovirus
  season from the four different models. Aggregated predictions refer to the final size by age group
  (Figure~\ref{fig:BNV_LTF_size}).}
\label{tab:BNV_LTF_DSS}
\begin{tabular}{r|rr|rr}
& \multicolumn{2}{|c|}{Weekly} & \multicolumn{2}{|c}{Aggregated} \\
 & mDSS & logDS & mDSS & logDS \\ 
  \midrule
reciprocal & 1.539 & 1.067 & 4.098 & 3.508 \\ 
  homogeneous & 1.564 & 1.096 & 4.205 & 3.393 \\ 
  no mixing & 1.521 & 1.078 & 4.066 & 3.742 \\ 
  power-adjusted & 1.527 & 1.065 & 4.071 & 3.638 \\ 
  
\end{tabular}
\end{table}

\section{Discussion}

We have provided a non-comprehensive review of the literature on
forecasting epidemics. We have focussed on statistical methods to
quantify the accuracy of predictions, distinguishing between point and
probabilistic forecasts. Two applications show how the different
techniques can be applied to uni- and multivariate forecasts.

Inspired by similar techniques in weather forecasting and other areas
of science, recent work in this area has focussed on model averaging
and stacking in order to improve the predictive quality of single
model forecasts \citep{RayReich2018}. Perhaps the biggest challenge to
epidemic forecasting is the incorporation of reporting delays and
underreporting, as described in Chapter V.3 of this handbook. \todo{check reference}
A rigorous analysis requires surveillance and internet search data
archived in a way where you can actually see what was available at a
given time \citep{McIver2014}. Such real-time epidemiological data should become
the standard in surveillance systems, facilitating the development of
novel forecasting techniques.


\section*{Acknowledgments}
We are grateful to the Swiss Federal Office of Public Health (BAG) for
access to the data on influenza in Switzerland. We thank Nicholas
Reich for helpful comments on a previous version of this chapter.

\putbib[references]
